\documentclass[useAMS,usenatbib] {mn2e}

\usepackage{times}
\usepackage{graphics,epsfig}
\usepackage{graphicx}
\usepackage{amsmath}
\usepackage{amssymb}
\usepackage{comment}
\usepackage{supertabular}

\newcommand{\kms}{kms$^{-1}$}

\newcommand{\acc}{\ensuremath{\rm atoms~cm^{-2}}}

\newcommand{\HI}{H{\sc i}}

\title {Modeling \HI distribution and kinematics in the edge-on dwarf irregular galaxy KK250}
\author [Patra et al.]{
	Narendra Nath Patra$^{1}$ \thanks {E-mail: narendra@ncra.tifr.res.in},
	Arunima Banerjee$^{1}$  \thanks {E-mail: arunima@ncra.tifr.res.in} 
	Jayaram N. Chengalur$^{1}$ \thanks {E-mail: chengalu@ncra.tifr.res.in}, 
	Ayesha Begum$^{2}$  \thanks {E-mail: ayesha@iiserb.ac.in} \\ 
	$^{1}$ NCRA-TIFR, Post Bag 3, Ganeshkhind, Pune 411 007, India \\
	$^{2}$ IISER-Bhopal,ITI Campus (Gas Rahat)Building,Govindpura, Bhopal - 23, India.
}
\date {}
\begin {document}
\maketitle
\label{firstpage}

\begin{abstract}

We model the observed vertical distribution of the neutral hydrogen (\HI) in
the faint (M$_{\rm B} \sim -13.7$~mag) edge-on dwarf irregular galaxy KK250.
Our model assumes that the galaxy consists of axi-symmetric, co-planar gas and
stellar discs in the external force-field of a spherical dark matter halo, and in vertical hydrostatic equilibrium. The velocity dispersion of the gas is left as a free parameter in the model. Our best fit
model is able to reproduce the observed vertical distribution of the  \HI\ gas,
as well as the observed velocity profiles. The best fit model has a large
velocity dispersion ($\sim 22$~\kms) at the centre of the galaxy, which falls
to a value of $\sim 8$~\kms\ by a galacto-centric radius of 1~kpc, which is
similar to both the scale-length of the stellar disc, as well as the angular
resolution of the data along the radial direction. Similarly we find that the
thickness of the \HI\ disc is also minimum at $\sim 1$kpc, and increases by about a
factor of $\sim 2$ as one goes to the centre of the galaxy or out to $\sim
3$kpc. The  minimum intrinsic HWHM of the \HI\ vertical distribution in KK250
is  $\sim 350$pc. For comparison the HWHM of the vertical distribution of the
\HI\  in the solar neighbourhood is $\sim 70-140$~pc.  Our results are hence
consistent with other observations which indicate that dwarf galaxies have
significantly puffier gas discs than spirals.
 
\end{abstract}


\section{Introduction}

\noindent  The equilibrium vertical gas distribution in a galaxy is determined by the balance between the vertical pressure gradients and the net gravitational potential of the galaxy. Measurements of the \HI\ scale-heights of nearby edge-on disc galaxies hence allow one to infer the vertical gravitational field. Further, since the \HI\ disc generally extends to galacto-centric radii where the gravity due to stars is negligible, both the flaring of the \HI\ disc as well as its rotational velocity can be used as diagnostic probes of the underlying potential of the dark matter halo \citep[see e.g.][]{olling95,becquaert97,narayan05,banerjee08,banerjee10,banerjee11}.
\noindent While modelling of the vertical distribution of the gas can be used to infer the properties of the dark matter distribution, comparison of models with observations is complicated by the following problem. The \HI\ disc scale-height can be constrained from  observations of edge on galaxies. In such a situation it is however difficult to measure the gas velocity dispersion ${\sigma}_{g}$, which is a critical parameter for the model.  \citet{banerjee11} show that the uncertainty in ${\sigma}_{g}$ is an important contributor to the uncertainty in the scale-height computed from the model. The calculated \HI\ scale-height is highly sensitive to the assumed value of ${\sigma}_{g}$; a 10\% change in ${\sigma}_{g}$ can lead to $\sim$ 15\% change in the calculated \HI\ scale-height.  While $\sigma_g$ can be measured for face on galaxies, in such cases it is diffuclt to directly measure the scale-height. The fact that it is difficult to simultaneously measure both the scale-height and the velocity dispersion makes it difficult to build completely data driven models of the vertical distribution of the \HI.

\noindent In this paper we use GMRT observations of the dwarf irregular galaxy KK250 to model the scale-height of the \HI\ distribution. The dark matter halo is assumed to be spherically symmetric and its parameters are fixed from modelling of
the observed rotation curve. Unlike the earlier studies, we do not assume a fixed value for the gas velocity dispersion, but instead  allow it to be a free parameter in the fitting.  We pick the model that best fits the observed vertical \HI\ distribution.
Although the observed velocity widths were not used as an input to the model, we also compare the observed line of sight profiles with the model ones, to check for consistency.  The rest of this paper is arranged as follows. In \S 2, we present the \HI\ data, in \S 3 the modelling procedure. \S 4 contains a comparison of the model with the observational data, and a discussion of the results.

\section{\HI\ observations and analysis}

\begin{figure*}
\begin{center}
\begin{tabular}{cc}
\resizebox{85mm}{!}{\includegraphics{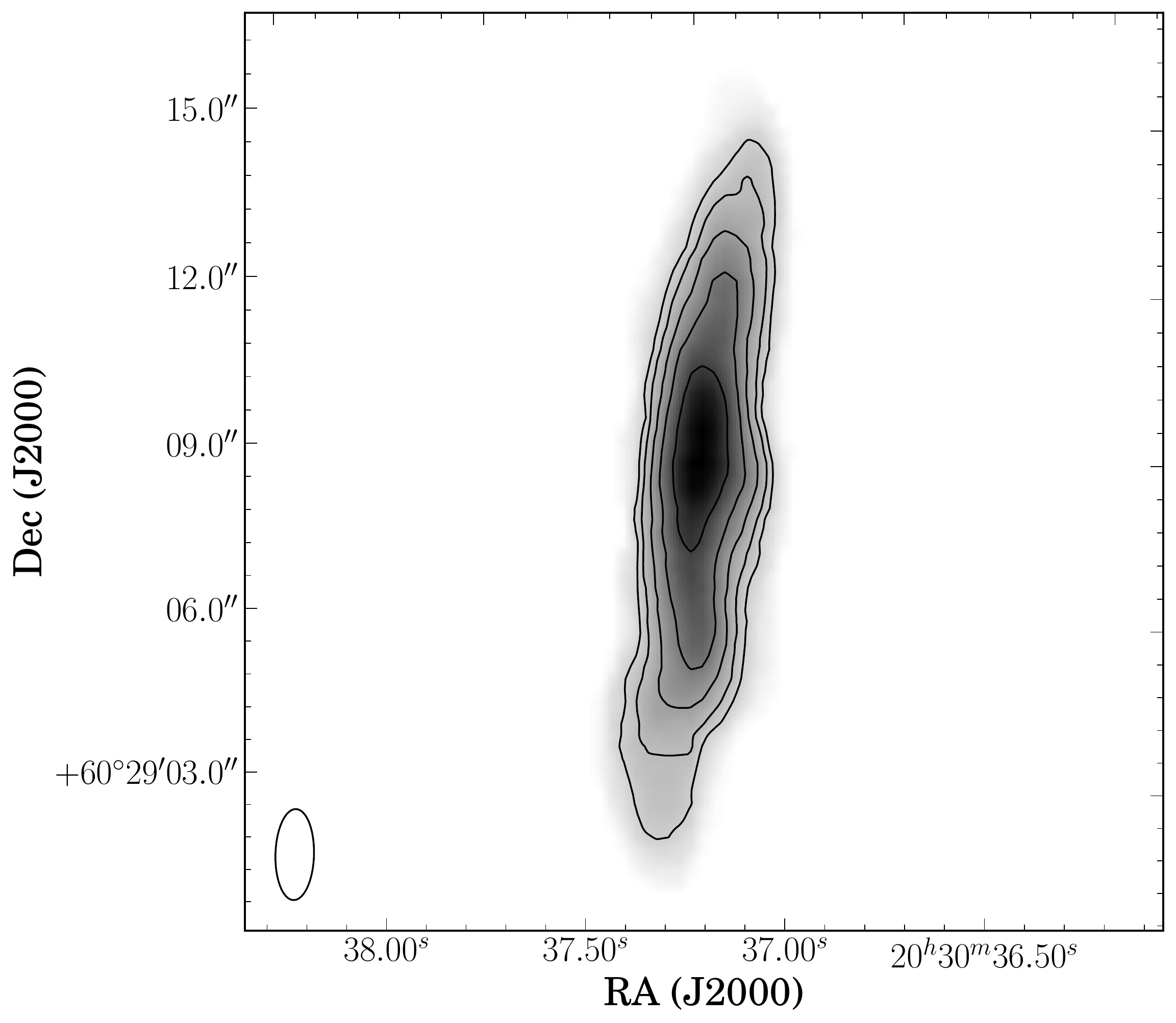}} &
\resizebox{85mm}{!}{\includegraphics{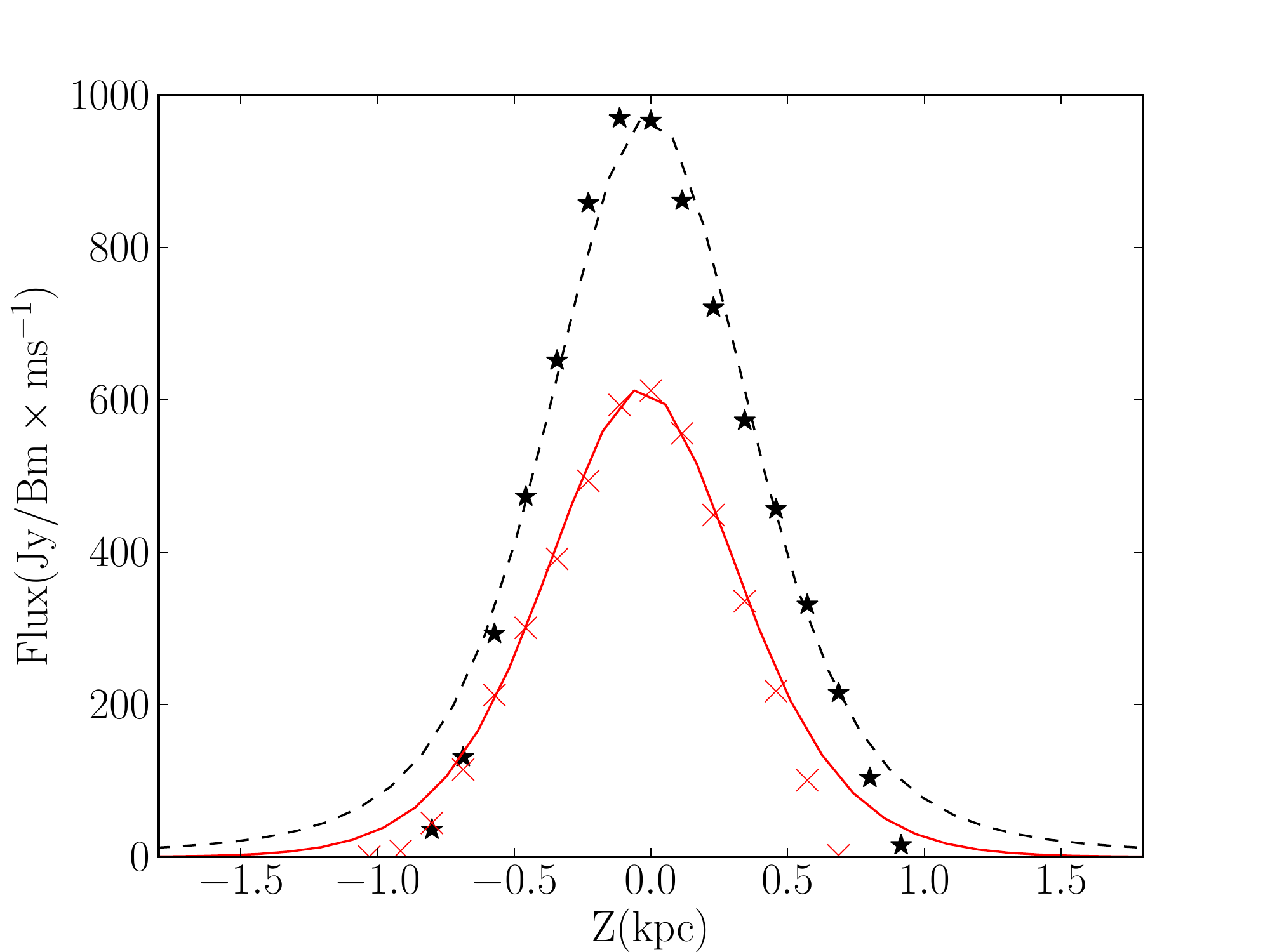}} \\
\end{tabular}
\end{center}
\caption{{\bf Left Panel} GMRT \HI\ 21 cm radio synthesis image of the dwarf irregular galaxy KK250. The image has been made using an elliptical beam with FWHM $32^{\prime \prime} \times 13^{\prime \prime}$. 
The contour levels are $(1, 1.4, 2, 2.8, ...)~ 5 \times 10^{20}$ \acc. See the text for more  details.
{\bf Right Panel} This shows the intensity profiles along two vertical cuts through the disc of KK250. The stars are at a galacto-centric distance of 0.1 kpc, while the crosses are at a galacto-centric distance of -1.9~kpc. The dashed and solid lines are the corresponding model profiles at these two locations.
}
\label{fig:mom0}
\end{figure*}

\noindent KK250 was observed as part of the Faint Irregular GMRT Galaxy Survey \citep[FIGGS;][]{begum08}. Throughout this paper we assume a distance of 5.9~ Mpc to the galaxy \citep{kar13}. At this distance, an angular separation of 1$^{'}$ corresponds to a linear separation of 1.7~kpc. The observations, data analysis as well as the rotation curve and mass decomposition for this galaxy are presented in \citet{begum04}. The galaxy is close to edge-on, meaning that a circular beam gives far fewer independent measurements along the minor axis than the major axis. In this paper we are interested in modeling the scale-height of the \HI. We would hence like to have as good a resolution as  possible  along the minor axis. The hybrid configuration of the GMRT \citep{swarup91} allows one to use a single GMRT observation to make images at a range of angular resolutions (varying from $\sim 40^{''}$ for data from the central square antennas alone to $\sim 2^{''}$ for data from the full array). However, the signal to noise ratio at the higher resolutions is generally modest. To get an optimum between the angular resolution along the minor axis and the signal to noise ratio, we choose to image the galaxy with an elliptical beam, with the major axis of the ellipse aligned with the major axis of the galaxy. We used the task UVSRT in AIPS to rotate the visibility data so that the major and minor axis of the beam are aligned with the galaxy axes. The desired ellipticity of the beam can then be obtained using the UVTAPER parameters to the AIPS task IMAGR. For the analysis below, the images were made with a synthesized beam having full width half maximum (FWHM) of 32$^{''}~\times$~13$^{''}$ (or 0.91~kpc $\times$ 0.37~kpc). The resulting image is shown in Fig.~\ref{fig:mom0}, along with the \HI\ intensity profile at two different galacto-centric radii.

\section{Modeling the galactic disc}

\subsection{Formulation of the Equations}

\noindent Our model for the galaxy is based on similar modeling of edge-on galaxies done earlier by \citep{narayan02,banerjee11}. Details of the model can be found in those papers, and is also briefly summarized below. The discs of the galaxy are modeled as a 2-component system of gravitationally-coupled stars and gas in the force field of a fixed dark matter halo. We assume that each of the two components is present in a form of a disc, the discs being concentric, and coplanar. We further assume the discs to be axi-symmetric and the halo to be spherically symmetric.  The joint Poisson's equation for the disc plus dark matter halo in cylindrical polar coordinates ($R$, $\phi$, $z$) therefore reduces to 

\begin{equation}
\frac{1}{R} \frac{\partial }{\partial R} \left( R \frac{\partial \Phi_{total}}{\partial R} \right) + \frac{\partial^2 \Phi_{total}}{\partial z^2} = 4 \pi G \left( \sum_{i=1}^{2} \rho_{i} + \rho_{h} \right)
\end{equation}

\noindent where $\Phi_{total}$ is the total potential due to all the components and dark matter halo. $\rho_i$ with $i = s,g$  denotes the mass density of stars and gas respectively. (Following \citet{begum04} we correct the observed \HI\ gas density by a factor of 1.4 to account for the presense of Helium.) $\rho_h$ denotes the mass density of the dark matter halo. In their modeling of the rotation curve of this galaxy, \citet{begum04} found that the dark matter halo is well fit by a pseudo-isothermal halo, and not well fit by an NFW \citep{navarro96} type halo. We hence take the density distribution in the dark matter halo to be  pseudo-isothermal, i.e. with a density profile given by 

\begin{equation}
\rho_h(R, z) = \frac{\rho_0}{1 + \frac{R^2 + z^2}{R_c^2}}
\end{equation}

\noindent {where $\rho_0$ and $R_c$ are the central core density and core radius respectively \citep{binney07}.}\\ 

\noindent The equation of hydrostatic equilibrium in the $z$ direction for each of the two disc components, is given by

\begin{equation}
\frac{\partial }{\partial z} \left(\rho_i {\langle {\sigma}_z^2 \rangle}_i \right) + \rho_i \frac{\partial \Phi_{total}}{\partial z} = 0
\end{equation}

\noindent where ${\langle {\sigma}_z \rangle}_i$ is the vertical velocity dispersion of the $i^{th}$ component ($i=s,g$), again an input parameter . \\

\noindent Eliminating ${\Phi}_{total}$ between Equation (1) and (3) we get 

\begin{equation}
\begin{split}
{\langle {\sigma}_z^2 \rangle}_i \frac{\partial}{\partial z} \left( \frac{1}{\rho_i} \frac{\partial \rho_i}{\partial z} \right) &= 
-4 \pi G \left( \rho_s + \rho_{g} + \rho_h \right)\\ 
&+ \frac{1}{R} \frac{\partial}{\partial R} \left( R \frac{\partial \phi_{total}}{\partial R} \right)
\end{split}
\end{equation}

\noindent Equation (4) can be further simplified using the fact

\begin{equation}
{\left( R \frac{\partial \phi_{total}}{\partial R} \right)}_{R,z} = {(v_{rot}^2)}_{R,z}
\end{equation}

\noindent Assuming that the vertical gradient in ${(v_{rot})}_{R,z}$ is small, we approximate ${(v_{rot})}_{R,z}$ by the observed rotation curve $v_{rot}$, which is a function of $R$ alone. Thus Equation (4) reduces to 
 
\begin{equation}
\begin{split}
{\langle {\sigma}_z^2 \rangle}_i \frac{\partial}{\partial z} \left( \frac{1}{\rho_i} \frac{\partial \rho_i}{\partial z} \right) &= 
-4 \pi G \left( \rho_s + \rho_{g} + \rho_h \right)\\ 
&+ \frac{1}{R} \frac{\partial}{\partial R} \left( v_{rot}^2 \right)
\end{split}
\end{equation}

\noindent Therefore Equation (6) now represents two coupled, second-order ordinary differential equations in the variables ${\rho}_s$ and ${\rho}_{g}$. The solution of Equation (6) at a given $R$ gives ${\rho}_s$ and ${\rho}_{g}$ as a function of $z$ at that $R$. Thus solving Equation (6) at all $R$ gives the model three dimensional density distribution of the stars and gas in the galaxy.

\subsection{Input Parameters}

\noindent As mentioned above, \citet{begum04} present mass models for this galaxy, based on the \HI\ rotation curve. We take the parameters from the mass model of \citet{begum04} as inputs to the current model. Specifically, the stellar surface density $\sum_*(R)$ is taken to be an exponential with a scale-length $R_D$ $\sim \ 1.1$~kpc with the total stellar mass $3.4 \times 10^7 M_{\odot}$.  The \HI\ surface density $\sum_{HI}$ is modeled as a double Gaussian and given by

\begin{equation}
{\Sigma}_{HI} = {\Sigma}_{01} e^{-r^2/2r_{01}^2} + {\Sigma}_{02} e^{-r^2/2r_{02}^2}
\end{equation}
with $r_{01} = 2.11 \pm 0.05$~kpc, $r_{02} = 0.41 \pm 0.02$~kpc, ${\Sigma}_{01} = 5.0 \pm 0.1 \ M_{\odot}$pc$^{-2}$ and ${\Sigma}_{02} = 4.2 \pm 0.2 \ M_{\odot}$pc$^{-2}$. As mentioned above, the \HI\ surface density is scaled to account for the presense of helium to get the total gas density. The dark matter halo is modelled as  a pseudo-isothermal halo (Eqn. 2) with $\rho_0 = 37.6 \times 10^{-3} \ M_{\odot}$pc$^{-3}$  and $R_c = 1.5$~kpc.

\noindent The remaining parameters that need to be specified are the vertical velocity dispersions in the stellar and gaseous discs. These cannot be directly  measured from the data for edge-on galaxies. Banerjee \& Jog (2011) show that while modelling the structure of the gaseous disc, the accuracy of the assumed value of stellar velocity dispersion ${\sigma}_{z,s}$  has a marginal effect on the calculated 
\HI\ scale-height. We hence assume the stellar dispersion ($\sigma_{z,s}$) to be the same as obtained analytically for that of an isothermal stellar disc in vertical hydrostatic equilibrium in the net potential of the stars and the dark matter halo \citep{leroy08}. The assumed value of \HI\ velocity dispersion, ${\sigma}_{g}$ however strongly regulates the calculated \HI\ scale-height \citep{banerjee11}. Therefore using an accurate value of ${\sigma}_{g}$ is fundamental to  obtaining the correct value of the \HI\ scale-height. Unlike earlier studies, we have not assumed an ad-hoc value for ${\sigma}_{g}$ for our calculations. Instead, as described below, we have treated ${\sigma}_{g}$ as a free parameter and constrained it using the observed data. \\

\subsection{Solution of the Equations}

We solve the two coupled, second-order ordinary differential equations given by Equation (6) in an iterative manner using a Eighth-Order 
Runge Kutta method as implemented in the scipy package.

\begin{equation}
\left( \rho_i \right)_{z = 0} = \rho_{i,0} \ \ \ \ {\rm and} \ \ \ \frac{d \rho_i}{dz} = 0
\end{equation}
  
The observations constrain the observed surface density ${\Sigma}_i$ and not the midplane density $\rho_{i,0}$. We hence take a trial value of $\rho_{i,0}$, and iterate until the model matches the observed $\Sigma_i$.  We are able to measure the \HI\ vertical distribution only within a projected galacto-centric radius of 3~kpc, beyond that the signal to noise ratio is too poor to allow a realiable estimate to be made. Accordingly we solve Equation (6) for 0 $<$ $R$ $<$ 3 kpc at steps of 100 pc in the radial direction, and an adaptive (scale-height sensitive) resolution along the vertical axis. The resolution along the vertical direction was always better than 10~pc. The solution at the intermediate radii is estimated by interpolation.  Since $v_{rot}$ is fixed, Equation(6) can be solved independently for each value of the radius $R$. Instead of assuming a fixed value of $\sigma_{g}$ at each $R$, we treat it as a free parameter satisfying the following two constraints: (i) 1 $<$ ${\sigma}_{g}$ $<$ 24~\kms (ii) ${\sigma}_{g}$ does not increase with $R$. The upper limit of $\sim 24$~\kms was set by increasing the upper limit to the range of $\sigma_{g}$ being searched until we found a value that bracketed the best fit $\sigma_{g}$. We then explored the entire available parameter space in velocity dispersion using a grid with  cell size in velocity of 2~\kms and radial cell size of 500~pc. In practice we generated a set of solutions for $\rho_i(R,z)$; each solution was for a fixed value of $\sigma_{g}$, with the $\sigma_{g}$ being varied in steps of 2~\kms between the different solutions. Combinations of these solutions were then made for all possible $\sigma_{g}$ profiles that satisfy the constraints listed above. Since, as noted earlier, Eqn(6) needs to be solved only as a function of $z$, combining the solutions is equivalent to solving the equation for each possible $\sigma_{g}$ profile. From the $\rho_g(R,z)$ and the rotation curve we constructed a 3D model of the galaxy. This was then projected to the adopted  morphological inclination ($i$ = $85^o$, see the discussion below), and convolved with the telescope beam to produce a model data cube.

\citet{begum04} estimated the morphological inclination to be $i$ = $80^o \pm 4^o$. A value of $80^o$ for the inclination requires an extremely  large velocity dispersion ($\sigma_g \sim 40$ \kms) at the centre to reproduce the observed morphology. In deriving the inclination, \citet{begum04} assumed the intrinsic axial ratio to be $q_0$ = 0.2, however \citet{roychowdhury13} show that dwarf galaxies tend to have somewhat larger intrinsic axial ratios. For galaxies with luminosity M$_{\rm B}$ between $-12.6 $ and $-14.8$ the mean axial ratio is $0.36$ with a scatter of $\sim 0.15$. A larger intrinsic axial ratio would lead to a higher value of the inclination. The  inclination adopted here ($i = 85^o$) is consistent with this expectation. We note however that there is some uncertainty in deriving inclinations of dwarf galaxies, and our adopted value is in some sense a best guess. 

  Since the \HI\ emission is optically thin, the integrated \HI\ flux image has folded into it the distribution of all of the gas along the given line of sight. This means that for an edge on galaxy there is no one to one mapping between the observed thickness of the \HI\ distribution at a given projected galacto-centric radius and the intrinsic \HI\ scale-height. In order to compare the model distribution with the observations, we hence generate a model integrated flux image from the model data cube and measure the FWHM of the \HI\ vertical distribution computed in exactly the same way as was done with the observed ``moment 0'' image. The $\sigma$ profile that produced the best fit (smallest $\chi^2$) fit to the observed data was picked as the best fit model. This is the most compute intesive step of the model fitting, and it was implemented using MPI based code and run on a 32 node IBM sandybridge cluster. A run using 400 processors takes about 8 hours. The best fit model had $\sigma_g$ tending to a constant value of $8$~\kms outside of $R \geq 1.5$~kpc, and rises steeply inside this. To resolve this better, a second set of models were produced, with $\sigma_{g}$ set to $8$~\kms outside $R = 1.5$~kpc, and with $\sigma$ varying from $7 - 24$~km/s inside. The cells of fixed $\sigma_g$ were reduced to 300~pc instead of the earlier value of 500~pc. The best fit of these models is then picked as our final model.

\section{Results and Discussion }

\begin{figure}
\begin{center}
\resizebox{85mm}{!}{\includegraphics{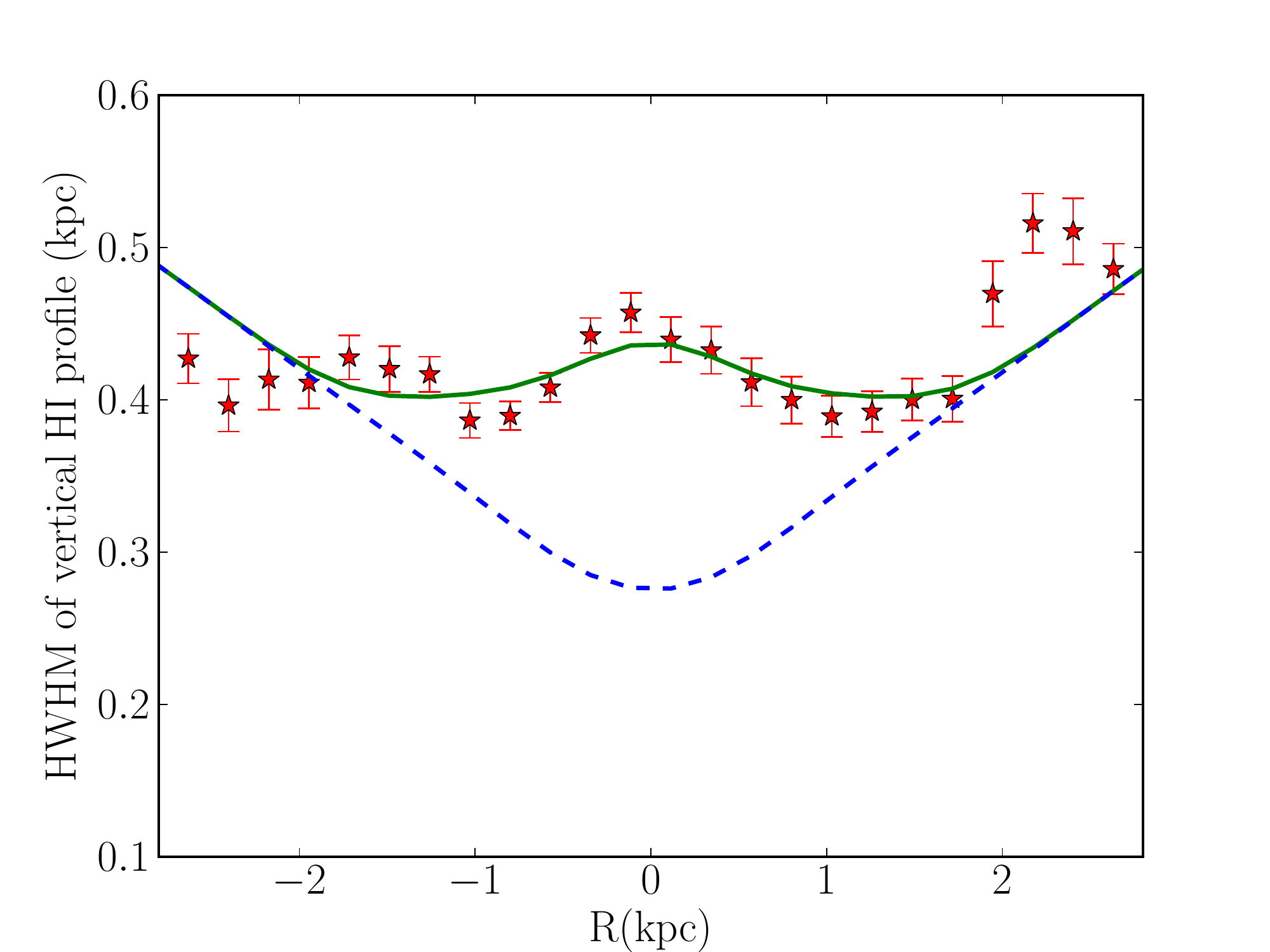}}
\end{center}
\caption{Plot of the obseved (points with error bars) half-width-at-half-maxima (HWHM) of the \HI\ vertical intensity profile of the edge-on dwarf irregular galaxy KK250 as a function of galactocentric radius $R$ overlaid  with the best fit model (solid line), as well as a model in which the gas velocity dispersoion, $\sigma_g$ is fixed at 8~\kms (blue dashed line).}
\label{fig:obs}
\end{figure}

 In Fig.~\ref{fig:obs}, we compare the half-width-at-half-maximum (HWHM) of the observed vertical \HI\ intensity distribution (points with error bars) with that from the best fit model (solid line). For comparison we also plot the HWHM profile for the constant $\sigma _g = 8$ \kms model (blue dashed line). As can be seen, the match between our model and the data is excellent inside a galacto-centric radius of 2~kpc. Between 2~kpc and 3~kpc the observed disc is slightly asymmetric. In this region one can see that the model represents some kind of average of the two halves. The model with constant $\sigma _g = 8$ \kms fails to produce the observed HWHM profile at the central region of the galaxy, though it merges with our model at $\sim R > 2$ kpc.

\begin{figure*}
\begin{center}
\begin{tabular}{cc}
\resizebox{85mm}{!}{\includegraphics{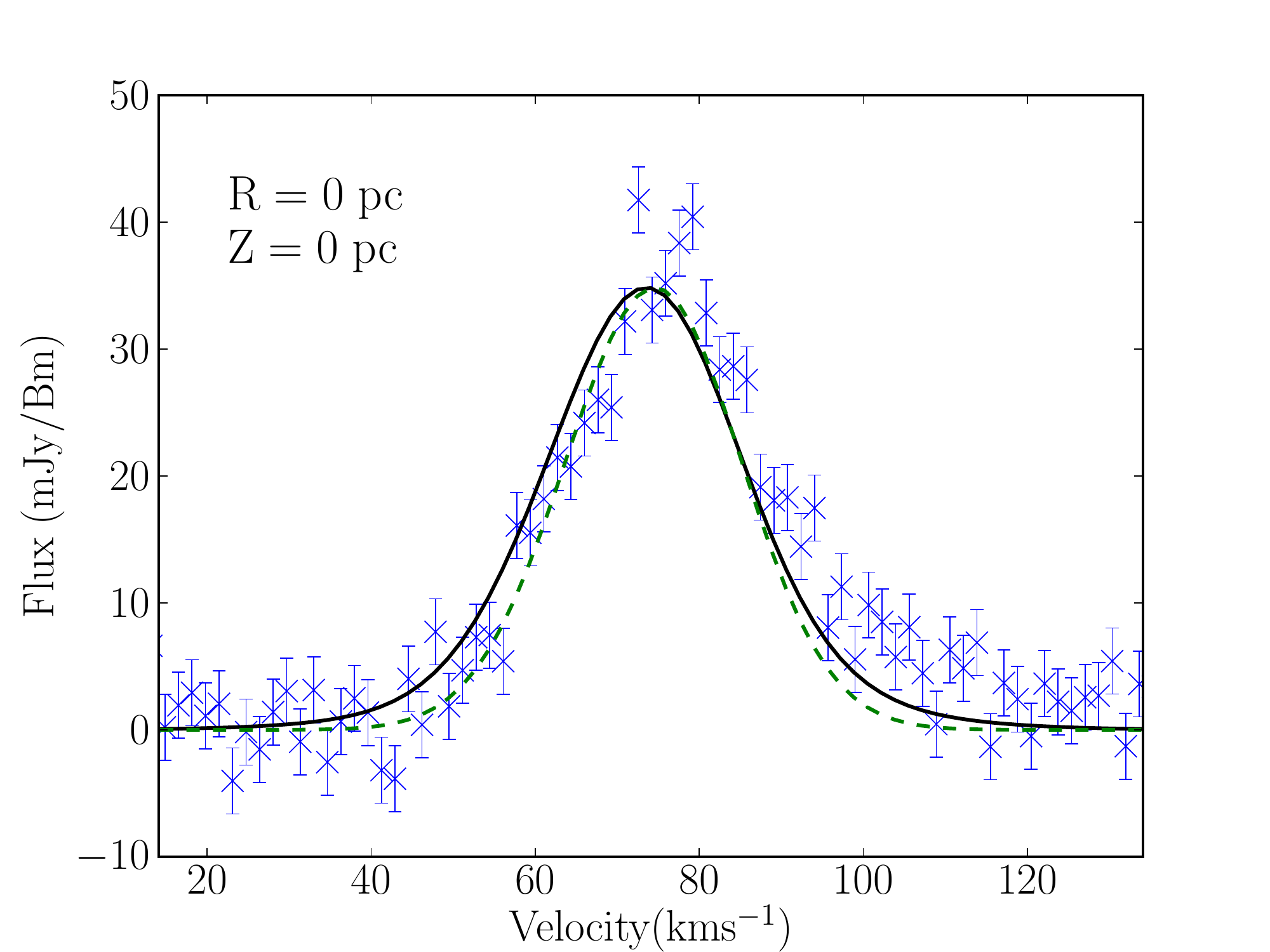}} &
\resizebox{85mm}{!}{\includegraphics{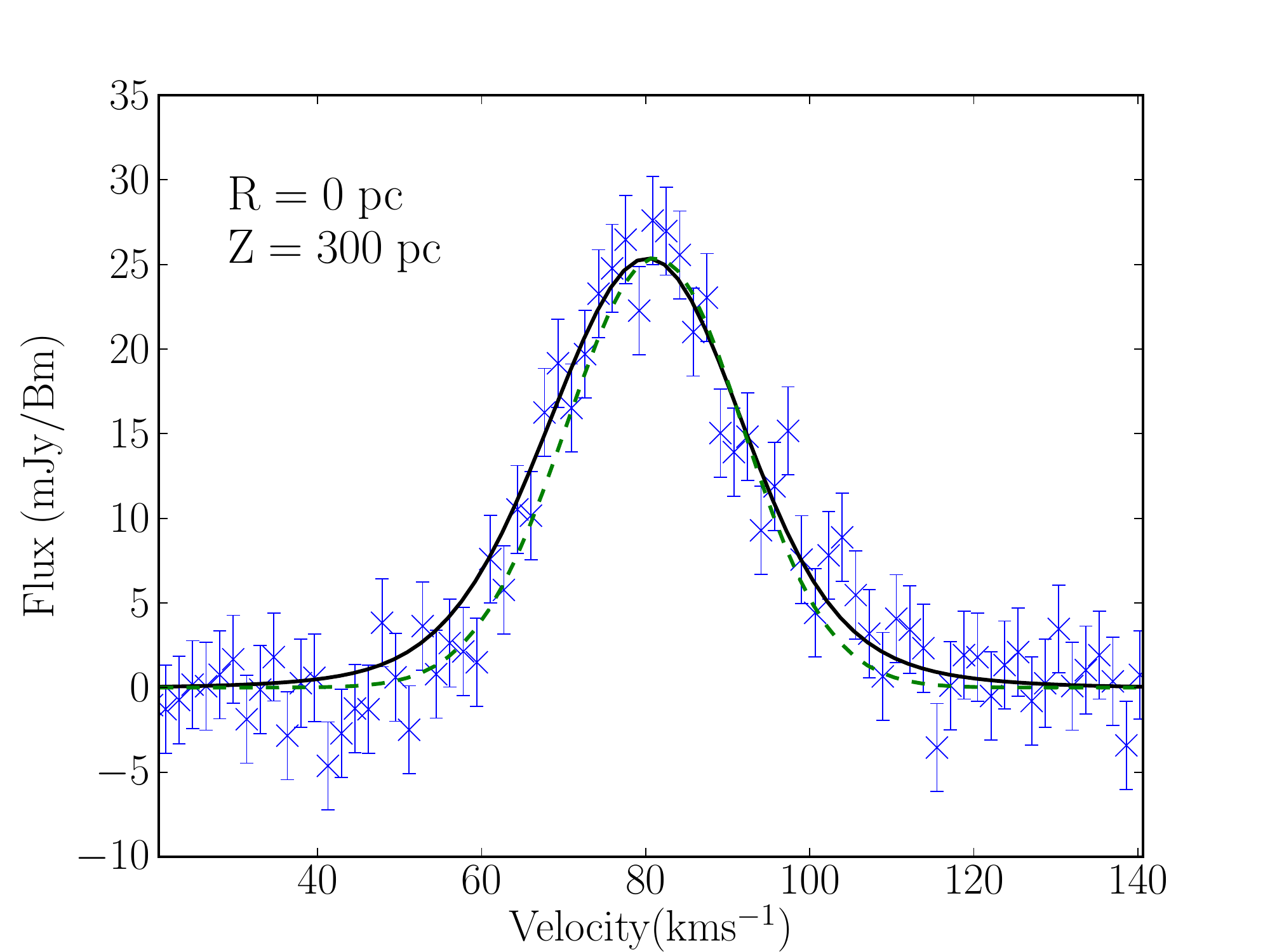}} \\
\resizebox{85mm}{!}{\includegraphics{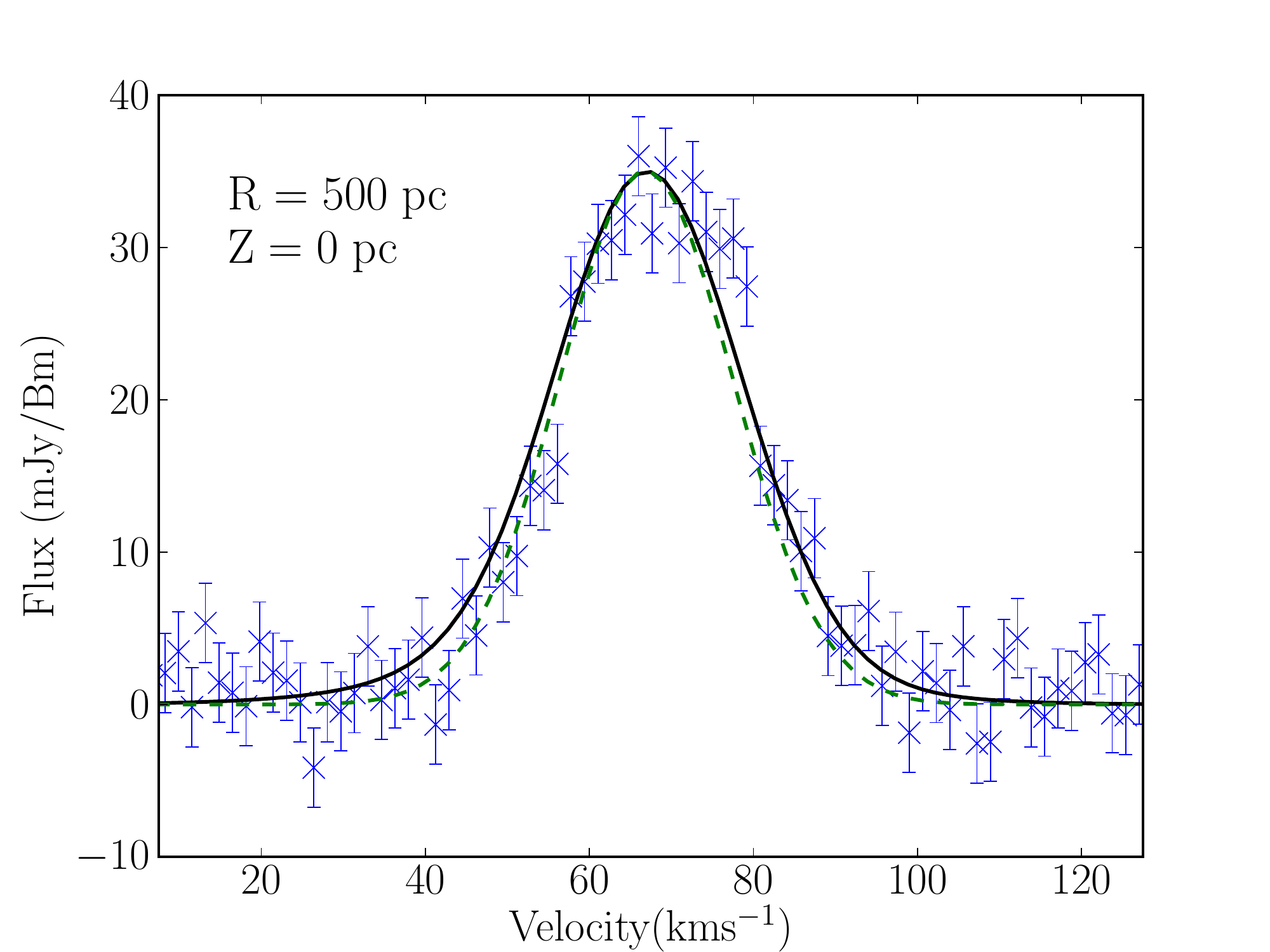}} &
\resizebox{85mm}{!}{\includegraphics{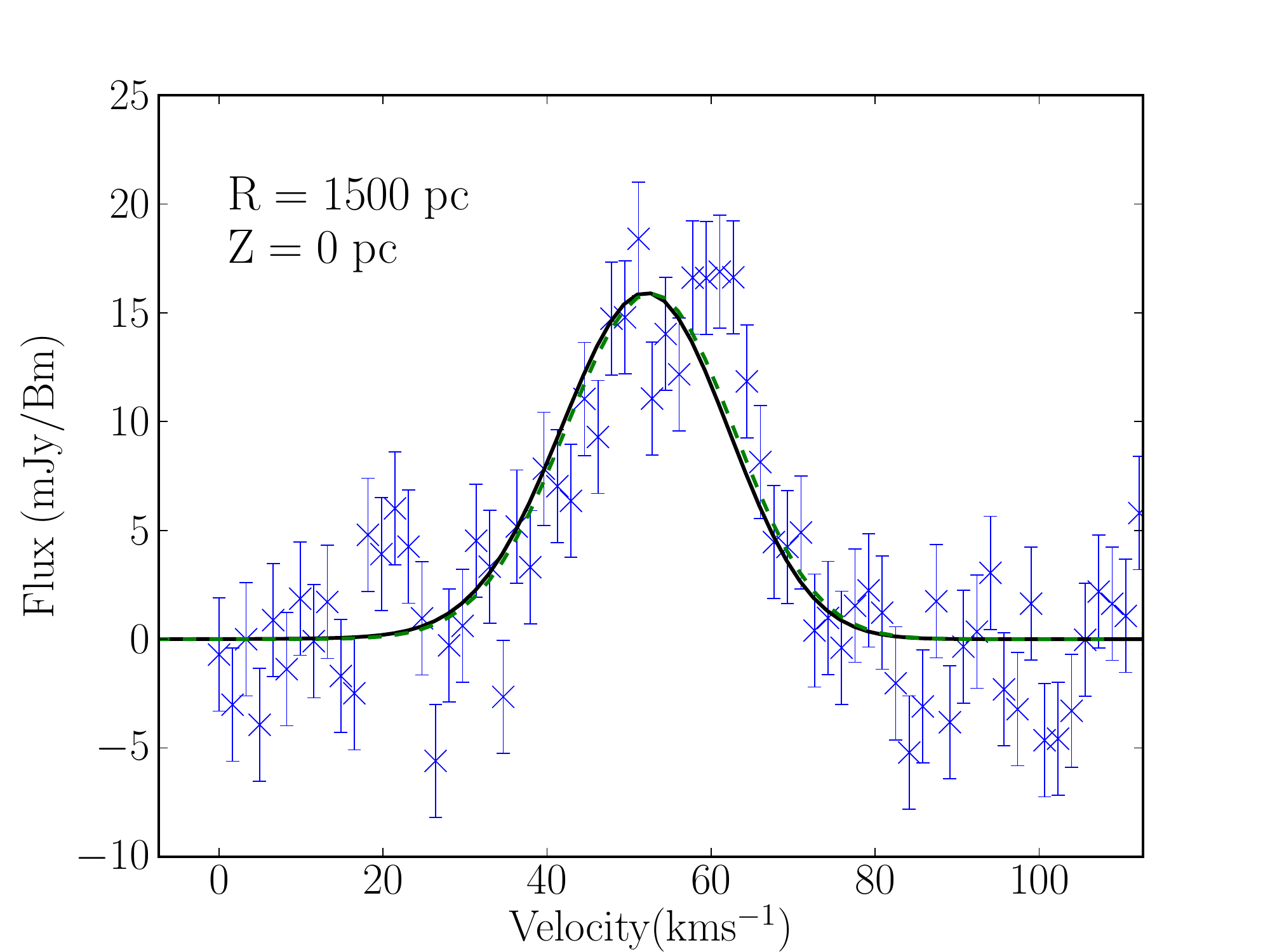}} \\
\end{tabular}
\end{center}
\caption{Observed versus model \HI\ spectra at different locations in the galaxy. The solid and dotted lines show the best fit model and a model with a constant velocity dispersion of 8~\kms respectively. The model was fit only using the morphological data and not the kinematical data. Comparison of the model and observed spectra is hence an independent cross check of the model.}
\label{fig:spec}
\end{figure*}

In Fig.~\ref{fig:spec} we compare the observed velocity profiles with the model profiles. Since the aim of this exercise is to see if the data provides some independent support for the existence of high velocity dispersion gas in the central regions of the galaxy, we have slightly shifted and scaled the model spectra so that they align exactly with the observed spectra. Recall that our model fitting used only the morphological data, this comparison provides an independent check. As can be seen from the figure, the velocity profiles in the central regions of the galaxy show an excess flux at $v\sim 100$ \kms as compared to the best fit model spectrum. This could perhaps be due to some low level non-circular motions which are not captured in our model. Several mechanism like feedback from star-formation, inflow or outflow of gas from the central region etc. could lead to this kind of non-circular motion. We also note in passing that the total amount of gas with high velocity dispersion is relatively small, and the existence of this gas is difficult to determine from the moment 2 maps that are often used as a measure of the velocity dispersion.

\begin{figure*}
\begin{center}
\begin{tabular}{cc}
\resizebox{85mm}{!}{\includegraphics{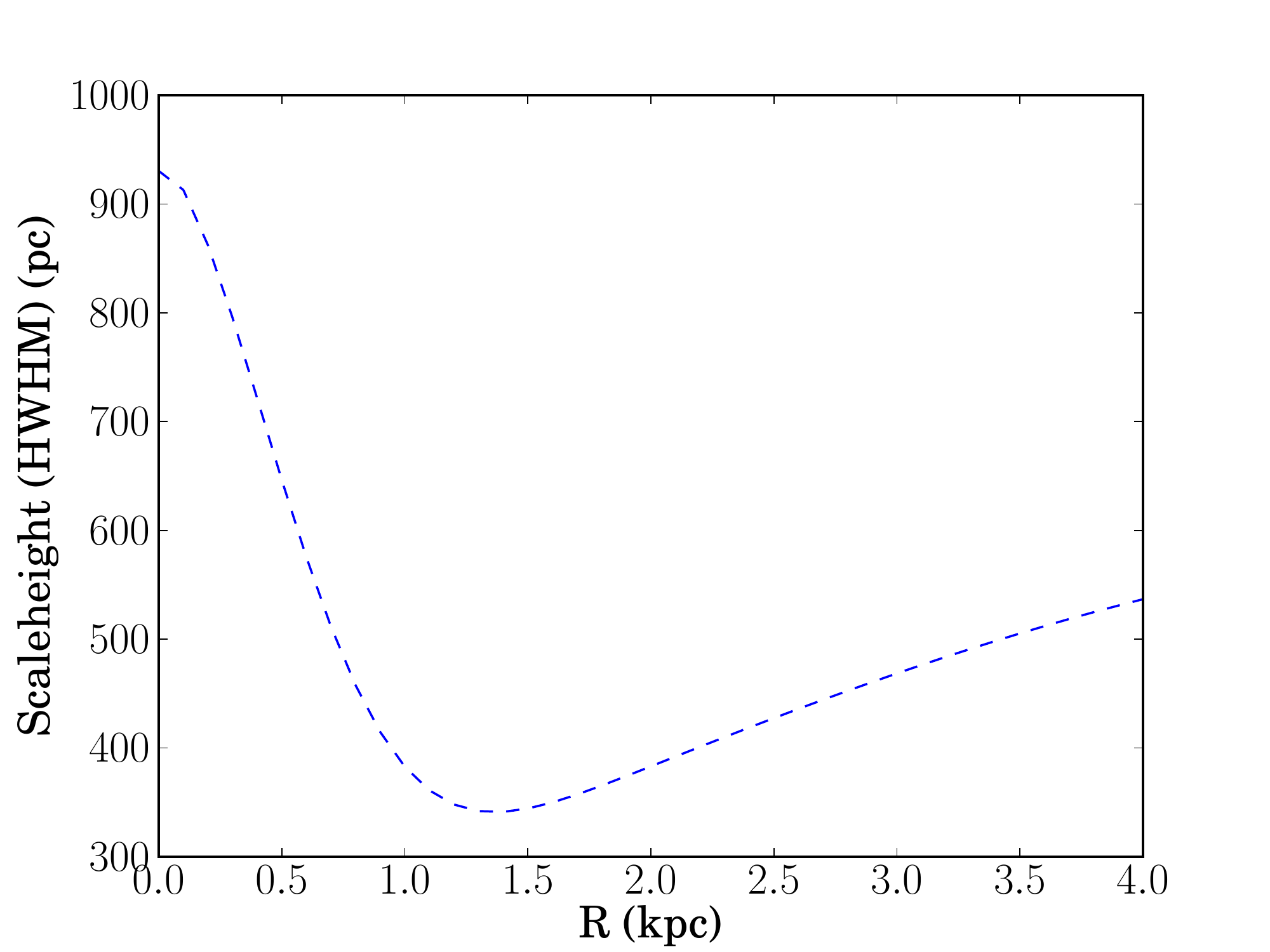}} &
\resizebox{85mm}{!}{\includegraphics{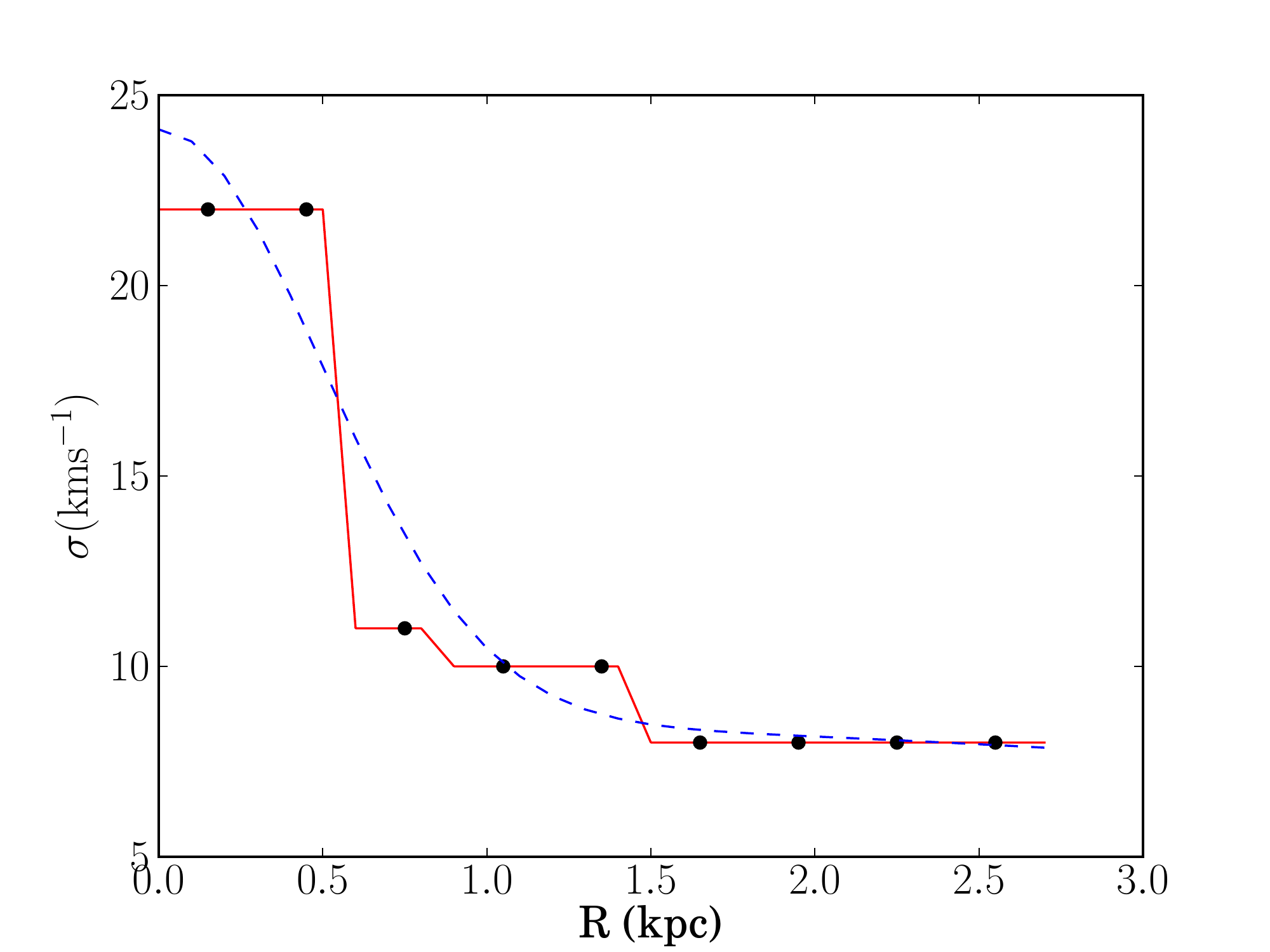}} \\
\end{tabular}
\end{center}
\caption{
{\bf Left Panel:} Plot of the calculated intrinsic scale-height (half-width-at-half-maxima) versus $R$ corresponding to the Gaussian fits to the best-fit $\sigma_{g}$ profile. 
\textbf{Right Panel:} Best-fit intrinsic \HI\ velocity dispersion ${\sigma}_{g}$ versus $R$ overlaid with a double gaussian fit to the same. The velocity dispersion  ${\sigma}_{g}$ takes a value of $\sim$ 22 \kms at the galaxy centre and falls steeply with increasing galacto-centric radius $R$ reaching to $\sim 8$ \kms beyond $R$ $\sim$ 1 kpc.}
\label{fig:scl_obs}
\end{figure*}

\noindent As discussed above, since the \HI\ is optically thin, the observed vertical distribution of the \HI\ reflects the scale-height distribution of all of the gas along the line of sight. If the scale-height does not vary with radius then the observed vertical distribution can be used to determine the intrinsic scale-height of the gas. If the scale-height varies with radius then determination of the true scale-height from the observed vertical distribution is not so straight forward. In the current case, since we have modelled the galaxy, we can use the model to determine how the scale-height varies with radius. In Fig.~\ref{fig:scl_obs} (Left Panel), we plot the calculated model scale-height (defined as the HWHM of the model \HI\ vertical density profile) versus $R$ for the best-fit ${\sigma}_{g}$ profile.  As can be seen the HWHM varies significantly with galacto-centric radius. At $R$ $\sim$ 0, the HWHM is $\sim$ 700~pc, it drops to a minimum value of $\sim$ 350 pc at $R$ = 1~kpc. This is  similar to both the scale-length of the stellar disc, as well as the spatial resolution of the data along the galaxy major axis. Beyond 1~kpc the gas disc flares by almost a factor of 2 by $R$ = 4 kpc. \HI\ flaring in the outer parts of galaxies is a common phenomenon and can be attributed to the decreasing importance of the self gravity of the disc components \citep[see for example the study by][]{banerjee11}. However the large HWHM seen in the central regions of the galaxy is unusual. To understand this better, it is useful to look at the radial variation of the velocity dispersion in our best fit model. In Fig.~4 (Right Panel), we plot our best-fit ${\sigma}_{g}$ versus $R$ along with a double-gaussian fit to the same. The velocity dispersion at the center of the galaxy is $\sim 22$~\kms and it falls steeply to a value of 8 \kms by $R$ $\sim$~1~kpc, which is equal to one exponential stellar disc scale-length $R_D$ . The earlier studies  \citep{banerjee11} of the scale-height in dwarf galaxies assumed a constant velocity dispersion of  $\sigma_{g} = 8$~\kms. In Figure~\ref{fig:obs} we show the HWHM derived from a model in which we have assumed $\sigma_{g} = 8$~\kms. As can be seen, this provides a poor fit to the data in the central regions of the galaxy. Clearly the high velocity dispersion in the best fit model is driven by the fact that the observed HWHM is roughly constant towards the center of the galaxy, while the vertical gravitational force gets larger as one approaches the galaxy center. The decrease in velocity dispersion as one goes from the center to the edge of the disc in KK250 is also consistent with earlier results that in most galaxies the velocity dispersion either remains constant or decreases with radius \citep{tamburro09}. Our inferred value of ${\sigma}_{g}$ of 22 \kms at the centre of galaxy is however somewhat high compared to the typical values obtained from other studies of dwarf  galaxies. For e.g. \citet{johnson12} find ${\sigma}_{g}$ $\sim$ 10 \kms for galaxies in the LITTLE THINGS survey and  \citet{stilp13} obtain values $\sim$ 6 - 12 \kms for the dwarf irregulars in the Local Group. We note however that apart from the center most region, the velocity dispersion in KK250 is similar to the values reported in these other works.

\noindent One of the input parameters to our model was the mass distribution as derived from the rotation curve by \citet{begum04}. In that paper the asymmetric drift correction (viz. the correction for the support against collapse provided by the pressure gradient in the gas) was  derived by assuming the velocity dispersion to have a constant value of 8 \kms. Since the value we get near the centre of the galaxy is somewhat higher, we recalculated the asymmetric drift-corrected rotation curve using the velocity dispersion as estimated here. The corrected curve lies   within the error bars of the original rotation curve of \citet{begum04}, which justifies the use of the mass decomposition derived in that paper. 

\noindent What is the origin of the velocity dispersion that we observe? The one dimension thermal velocity dispersion associated with 8000~K gas (i.e. a reasonable upper limit to the kinetic temperature of the Warm Neutral Medium) is $\sim 8$~\kms. So while it is possible that the velocity dispersion that we see in the outer parts of the galaxy are entirely due to thermal motions of the particles, in central regions there must be an extra contribution due to small 
scale bulk motions of the gas. Velocity spreads in excess of thermal in the neutral ISM are often attributed to turbulence. In the case of dwarf galaxies, further evindence for the exisitence of turbulence in the atomic ISM comes from studies of the power spectrum of the \HI\ intensity fluctuations \citep{begum06,dutta09,block10,dutta13}. Turbulence in the atomic gas in galaxies can arise from energy input from star formation and supernovae explosions \citep{maclow04},  magneto-rotational instabilities (MRI) in differentially rotating discs, \citep{sellwood99}, gravitational instabilities \citep{wada02}  and mass accretion \cite{klessen10}. KK250 has a linearly rising rotation curve, and hence the magneto-rotational instability does not appear to be relevant here. KK250 has a nearby companion dwarf galaxy, KK251, and tidal interaction could be a possible reason for the increased velocity dispersion. However, the \HI\ distributions of KK250 and KK251 show no obvious sign of interaction or disturbance, and  the rotation curves of the galaxies also appear fairly regular \citep{begum04}. It is hence appears that tidal interactions may not be playing an important role in this system. Which, if any, of the remaining mechanisms are responsible for driving turbulence in dwarf irregulars is currently unclear \citep[see for e.g.][]{stilp13}.\\ 

\noindent Regardless of the origin of the velocity dispersion, the fact that the estimated velocity dispersion in dwarfs is comparable to that in much larger spiral galaxies will have consequences for the structures of the gaseous discs in dwarfs. The larger ratio of ${\sigma}_{g}/v_{\rm rot}$  in dwarfs will lead to their gas discs being thicker that those of spiral galaxies. Indeed, the typical HWHM for the \HI\ that we determine for KK250 is significantly larger than the value of $\sim 70-140$pc determined for the inner disc of the Milky way \citep{malhotra95}.
 Similarly large scale-heights for dwarf galaxies have also been indicated by studies of the power spectrum of the \HI\ intensity fluctuations \citep{dutta09}. These results are based on the expectation that the slope of the power spectrum will change on scales comparable to the scale-height of the galaxy \citep{elmegreen01}. Finally, studies of the axial ratio distribution of the gas and stellar discs of dwarf galaxies \citep{roychowdhury10,roychowdhury13} do indeed show that their stellar and gaseous discs are significantly thicker than those of spirals. The  more diffuse gas distribution has several implications for galaxy evolution, including being likely to be one of the causes for the observed low efficiency of conversion of stars to gas in dwarf galaxies \citep{roychowdhury09,roychowdhury11}.

\section{Conclusion}

We use GMRT \HI\ 21cm observations of the dwarf irregular galaxy KK250 to model the vertical distribution of the atomic gas. Our model consists of a 2-component galactic disc model of gravitationally-coupled stars and gas in the force field of a dark matter halo. The dark matter halo parameters are derived from fitting to the observed rotation curve. We allow the velocity dispersion of the gas to be a free parameter in the model, and constrain it using the observed half width half maximum of the vertical distribution of the \HI. For the best fit model we also compare the observed line of sight velocity widths with the model ones, and find the two to be in good agreement. This provides a further consistency check to the derived velocity dispersion. From our best fit model, we find that the velocity dispersion is maximum at the center, but rapidly falls to a value of $\sim 8$~\kms at $R \sim 1$~kpc. This is similar to both the scale-length of the stellar disc, as well as the spatial resolution along the radial direction. The  intrinsic \HI\ scale-height also reaches its minimum value at $R \sim 1$~kpc. The scale-height that we measure for KK250 is significantly larger than the scale-height of the \HI\ in the central regions of the Milk way, consistent with earlier observations that dwarf galaxies have significantly thicker gas discs than typical spiral galaxies.

\end{document}